\documentclass[conference]{IEEEtran}
\ifCLASSINFOpdf
\else
\fi
\usepackage{epsfig,endnotes}
\usepackage[bookmarks=false]{hyperref}
\usepackage{url}
\usepackage{cite}      
\usepackage{microtype} 
\usepackage{cleveref}  
\renewcommand{\url}[1]{\href{#1}{#1}} 

\usepackage{amsmath,amssymb,amsthm}
\usepackage{xspace}

\usepackage[disable]{todonotes}
\newcommand{\mv}[1]{\todo[inline,color=green!30]{#1}}
\begin{document}
\begin{NoHyper}
%
\title{Revealing the Unseen: How to Expose Cloud Usage While Protecting User Privacy}

\author{\IEEEauthorblockN{Ata Turk}
\IEEEauthorblockA{Boston University\\
ataturk@bu.edu}
\and
\IEEEauthorblockN{Mayank Varia}
\IEEEauthorblockA{Boston University\\
varia@bu.edu}
\and
\IEEEauthorblockN{Georgios Kellaris}
\IEEEauthorblockA{Boston and Harvard Universities\\
kellaris@bu.edu}}


%


\maketitle

\begin{abstract}
Cloud users have little visibility into the performance characteristics and utilization of the physical machines underpinning the virtualized cloud resources they use. This uncertainty forces users and researchers to reverse engineer the inner workings of cloud systems in order to understand and optimize the conditions their applications operate. At Massachusetts Open Cloud (MOC),  as a public cloud operator, we'd like to expose the utilization of our physical infrastructure to stop this wasteful effort.  Mindful that such exposure can be used maliciously for gaining insight into other user’s workloads, in this position paper we argue for the need for an approach that balances openness of the cloud overall with privacy for each tenant inside of it. We believe that this approach can be instantiated via a novel combination of several security and privacy technologies. We discuss the potential benefits, implications of transparency for cloud systems and users, and technical challenges/possibilities.
\end{abstract}


%
\IEEEpeerreviewmaketitle

\section{Introduction}
\label{sec:intro}

Computation is rapidly migrating to the cloud due to its economies of scale and excellent network connectivity. Large companies operate thousands of VMs on the cloud~\cite{RightScale:2017}; government agencies, non-profit organizations, and research institutions adopt cloud based IT solutions~\cite{EducationCloud, GovernmentCloudUsage, AWSGovernmentCloud}, while advances in mobile and IoT solutions shift their computation towards the cloud~\cite{Botta:2016}. 

Application development in the cloud has different dynamics compared to traditional practices. The freedom to choose the size and type of instances that will operate applications brings along the problem of identifying/selecting the best instances for running the applications.
Here, `best' does not mean the fastest or most powerful instance, but rather the one whose specific configuration is best-suited for the intended application at the lowest cost.
Variability in cloud instance performances due to external interferences (e.g., noisy neighbors), especially in smaller sized cloud instances~\cite{delimitrou2016,noisyneighbor}, further complicate this selection process. 

Cloud application developers (e.g. Netflix) spend an enormous amount of effort and money to develop and execute software for instance benchmarking, monitoring, and performance evaluation \cite{NetflixNodeTuning}. This software allows the prospective tenants to select and operate the right set of cloud resources that will optimize the performance-to-cost-ratio of their applications. 

In order to aid cloud users in their endeavors to evaluate cloud instances, cloud providers (e.g. Amazon CloudWatch~\cite{AWSCloudWatch}) and various monitoring companies (e.g. Datadog~\cite{Datadog}, LogicMonitor~\cite{Logicmonitor}) monitor and expose application-level resource utilization and performance features. These systems can report application interactions with various services (e.g. number of EBS calls) and monitor performance based on user defined application performance metrics via instrumenting user applications. 

Instance selection may be performed by using rough generic rules, by sampling and benchmarking on small scale test cases and using that information for instance performance prediction~\cite{venkataraman2016}, or by all-out brute-force benchmarking on all possible instance types to identify the best fitting options~\cite{NetflixNodeTuning}. These solutions utilize the metrics collected from applications in order to perform instance selection. 

Unfortunately application-level performance metrics are indirect: Problems such as noisy neighbors (e.g. a neighbor causing many L2 cache misses), which are observed due to shared usage of underlying physical resources, manifest in the metrics after the problem impacts the performance of the application.    

\paragraph{Problem statement}
Current level of information provided by cloud vendors forces users to reverse engineer physical resource utilization metrics, which is a misguided and wasteful process. At Massachusetts Open Cloud, an academic public cloud, we would like to be better merchants and provide physical resource utilization measurements as a service (potentially even free of charge to differentiate our services) to our users. 

The potential benefits of exposing physical resource utilization are many-fold. For example, by exposing the number of L2 cache misses on the physical host where a cloud instance is running, we can enable better performance prediction for cache sensitive applications and trigger problem prevention mechanisms to ``kick-in.'' Similarly, exposing the ``current connectivity'' of physical hosts to services in the cloud (e.g. block storage solution) or to the outside world can be significantly useful during instance selection.  
\mv{Ata: At some point, it'd be great to say \emph{why} network connectivity info is useful. For instance, are there differences in intra-datacenter transit times for physical machines in different parts of the datacenter?}
Beyond its value to commercial application developers, publicly-accessible data on cloud utilization also provides immense value to researchers. First, for any experiment run on the cloud, utilization data can provide context for the results and facilitate replication and extension of the work. Second, utilization data can enable researchers to study the operation of a datacenter itself.

Unfortunately, the same data that can provide value for cloud tenants also have the potential to be exploited by malicious entities to gain insight into other users' workloads, reducing the security posture of the cloud by enabling hazardous activities such as co-location and side-channel attacks. The problem at hand is to come up with mechanisms that enables us to enjoy the benefits of information release while mitigating the  security implications as much as possible. 

\paragraph{Contributions}
In this paper we discuss the potential benefits and harms of releasing physical utilization data of public clouds in a publicly-accessible form. Next, we examine several technologies that address portions of the tension between openness and security. We propose that these technologies combine synergistically to cover each others' limitations and therefore offer a compelling solution that addresses the needs of cloud vendors and tenants alike. Finally, we pose a list of questions that we'd like to discuss with the PAIS community.
\section{Tradeoff: Transparency vs Security}

\begin{table*}[tbh]
\centering
\begin{tabular}{lll}
Type of user & Objective & Pertinent metric\\
\hline
Current or future tenant & Purchase/use instance with desirable
performance 
& Accuracy of provided data\\
Researcher using the cloud & Denote conditions of cloud at time of experiment & Reproducibility\\
Cloud researcher & Obtain aggregate cloud system statistics & Precision of analysis\\
Current tenant & Prevent co-location attacks & Probability of being pinpointed\\

\end{tabular}
\caption{Categories of cloud customers. We discuss the benefits to the first three categories of users in \Cref{benefits} and the potential security concerns to the fourth category in \Cref{harms}.}
\label{customers}
\end{table*}


We believe that increased transparency regarding cloud utilization will reduce user costs and increase the viability of using cloud services. Both of these features provide market incentives for a cloud provider to offer utilization metrics as a service. On the other hand, utilization data can expose activities of tenants on the cloud. More specifically, aggregate utilization data may be used to pinpoint a specific tenant within the cloud. An ideal solution needs to resolve the tension between transparency and user security favorably, maximizing transparency while minimizing security risks to tenants. 

Current cloud offerings do not investigate the trade-off between transparency and security. They simply offer no transparency and potentially maximum security. We dispute this decision. Since the importance of both objectives are subjective, we start our discussion by listing user classes that could benefit from cloud transparency and user classes that could be harmed by it.

\subsection{Beneficiaries of Cloud Transparency}
\label{benefits}

In this section, we distinguish between three distinct types of users who can benefit from the exposition of cloud physical resource utilization information in very different ways (see \Cref{customers}).


\paragraph{Current and future tenants}
Physical utilization data are of immense interest to tenants of the cloud. They know the CPU, memory, disk, and network performance requirements of their application better than anybody else. Physical utilization data can enable them to navigate the performance unpredictability of instances allocated to them. 

Currently, tenants spend extensive time and money on reverse engineering efforts to `vet' several virtualized resources and gauge their relative value toward the target application. A cloud provider who offers physical utilization as a service can free prospective tenants from this burden and permit them to focus on optimizing their own application rather than the operation of the cloud overall.

\paragraph{Scientific researchers}
Today, even though many research studies are conducted using cloud resources, repeatability of experiments conducted on the cloud are debatable. This variability concern when conducting experiments on the cloud forces researchers to repeat experiments many times incurring higher cost. Still the analysis made in many scenarios can be dependent on the state of the cloud at the assessment time (e.g., an experiment made on a busy working day or a working hour may not match an experiment conducted over the weekend.)

Currently it is not possible to exactly say under what conditions the experimental findings are gathered. If utilization information associated with the physical resources used in the experiments were available to researchers, these could be reported along with the experimental parameters to shed light to external effects that impact experimental observations.




\paragraph{Cloud designers and engineers}


One can argue that current ``black-box'' design of pubic clouds prevents cloud system innovation to happen in places other than the few big cloud vendors. Utilization data open up exciting opportunities for researchers who study the design of a cloud itself. For instance, utilization data can permit engineers to examine the impacts of different load distributions on cooling systems at scale without the expense of building a realistic datacenter and cloud to `test' theories. 

\subsection{Bearers of Potential Security Harms}
\label{harms}


Utilization data may be used to pinpoint a specific tenant within the cloud. Hence under a more transparent cloud model, current tenants have to bear the risks of identification. 
This is a critical risk for two reasons.
First, an attacker could passively use this information to monitor the tenant's use of the cloud (e.g., for one company to examine the popularity of a competitor).
Second, an attacker can actively spawn a co-located VM to invade the tenant's privacy by, e.g., 
determining the tenant's software~\cite{IrazoquiIES15,SuzakiIYA11,ZhangJRR14} or cryptographic keys~\cite{OsvikST06,InciGAES15,BengerPSY14}.

Below, we detail several existing methods that an attacker might be able to use to identify the tenant's physical location within the cloud.

\paragraph{Cache usage}

Perhaps the most common category of side-channel attacks on the cloud involves shared caches between tenants on the same physical machine. Attacks of this type permit an attacker to check whether a victim with predictable behavior (e.g., running a known web server) appears on a specific physical machine. More specifically, several papers have demonstrated the viability of \textsf{Prime+Probe} and \textsf{Flush+Reload} attacks on the cloud, whereby the attacker manipulates the cache with her own user-space process and then observes how her memory speeds are impacted by the target tenant's process \cite{RistenpartTSS09,InciGAES15,BengerPSY14}.

So far, many of these methods have required active effort by the attacker even to perform identification, much less to extract data from the tenant afterward. Furthermore, several countermeasures have been proposed that permit the tenant to leverage the cache side-channel for her own defensive purposes \cite{InciGES16,ZhangJOR11}; at a high level, they have the tenant monitor the cache herself to classify when its behavior is consistent with a \textsf{Prime+Probe} or \textsf{Flush+Reload} attack.

Unfortunately, our plan to have the cloud publish physical utilization (e.g. CPU, RAM) means that we have reduced the attacker's burden from an active role to a passive one: she receives the actual cache patterns that the side channel attacks attempt to extract! Furthermore, the defensive countermeasures are rendered meaningless as well. As a result, we must design a system that adequately protects tenants' sensitive cache information even while still providing the benefits described in \Cref{benefits}.

\paragraph{Network usage}

The tenant's network utilization may also be used to identify her. An attacker can influence the network latency or bandwidth available to the target tenant and observe the impact. For instance, imagine that the tenant uses the cloud to run a web server. Then, the attacker can probe the web server with a specific network flow and then fingerprint the tenant based on which network trace demonstrates the same pattern \cite{RistenpartTSS09,GongKB10,Bates2014}. If the attacker is already co-resident on the cloud, then she can verify the target's presence by flooding their shared link and observing a corresponding dip in the target's connectivity \cite{BlockN15,AgarwalMHHV16}.

Even with public network utilization data, the attacker still requires some sort of active posture to conduct any of the above attacks. But, this `active posture' may be as simple as making selective queries to a public-facing website. As a result, in order to realize our vision, we must adequately obscure or hide the tenant's current network consumption so as to thwart the attacks described above.

\paragraph{Differences over time}

The attacks described above share a common property: in principle, they can be executed simply by providing utilization data at a single point in time. However, our vision is even stronger than this: we wish to update the utilization data periodically.

Differencing attacks \cite{lucero11} allow attackers to infer the location of users based upon changes in utilization over time.
They operate by observing small changes between two similar states, in order to disclose an individual’s confidential data. 
For example, if a single service changes behavior, while the rest of the users continue to consume the same resources, an adversary can observe the location of the change, essentially locating the service.
We note that these changes can occur within the same machine or across physical machines on the cloud.

\section{A Path Forward}

In this section, we investigate methods to overcome the tradeoff between openness and security.

\mv{We can state that our proposed approach is just one way to solve the problem, and there may be others. The point is that the vision may be more generic than the proposed solution.}

A robust technique should allow the accurate monitoring of cloud resources, while ensuring the security of the cloud users. Concerning the latter, a secure approach must hide the location of each user by protecting against (i)~active attacks performed by adversaries that are also tenants, and (ii)~passive attacks performed by adversaries observing the published statistics. In the first part of this section, we investigate four security technologies that individually provide some (but not all!) of the necessary protections.

Then, we propose a path forward via a novel combination of the four security-enhancing technologies. At its core, our resolution to the openness-security tradeoff involves the mismatch in \emph{time}: attackers require information about the location of a victim's VM at the time of an attack, whereas honest beneficiaries of cloud utilization data are primarily concerned with the historical performance of a physical machine independent of the people who happen to be present on it at the moment.
Ergo, published metrics can balance between openness and security by providing a variable notion of accuracy that provides accurate utilization in the past but whose fidelity \emph{decays} as time moves toward the present.


\subsection{Existing Tools}

In this section, we describe four technologies that provide some protections for cloud tenants against identification attacks. For each technology, we provide a brief description of its operation, and then focus on its capabilities and weaknesses at addressing our specific security needs.

First, \emph{Secure Cloud Scheduling} (SCS) offers the promise of thwarting active attacks on the cloud \cite{AzarKM0S14}. It places VMs in a manner that is unpredictable to the adversary and that reduces its probability of successfully completing a
co-location attack. Although it offers high performance, it only ensures a weak form of security. In particular, its scheduling, while adversarially-controlled, is static. Hence, if an adversary ever manages to identify a target VM, she will know this information indefinitely.



Second, \emph{Moving Target Defense} (MTD) \cite{jajodia11} leverages continuous decision-making to benefit the defender. In the cloud setting, MTD allows tenants to migrate their VMs to new physical hosts within the cloud by means of a scheduler that remains securely outside of the attacker's view \cite{Valizadeh2016MarkovMO}. This movement increases the attacker's uncertainty and therefore increases her overall attack cost. Additionally, it mitigates the damage from prior revelations (overcoming SCS's defect), thereby reducing the attacker's window of opportunity.

However, MTD has two drawbacks. First, the migration of VMs is resource-intensive; if the cloud needs to perform this often at the datacenter scale, the cloud's overall efficiency will deteriorate. Second, MTD is feckless against problems that only require a passive adversary to exploit. Recall from \Cref{harms} that some previously-known attacks are susceptible to passive adversaries; more importantly, the public availability of utilization data transforms some active attacks into passive ones. In these cases, MTD allows invasive attacks for as long the user remains in the exposed location.

Third, \emph{Post-Compromise Security} (PCS) \cite{Cohn-GordonCG16} protects data and software in the present even if secret key material was compromised in the past. The dual to forward secrecy \cite{DiffieOW92,UngerDBFPG015}, PCS is the culmination of a long body of literature into authenticated key exchange \cite{CanettiK01,LaMacchiaLM07,StebilaU09,Krawczyk05}.
It is well-suited for combination with MTD because it can restore security guarantees against a tenant whose previous information has been compromised.
For this restoration to be possible, though, the adversary must lack the information needed to compromise the target in the present. So far, all of the technologies discussed fail to provide a distinction between past and present.

Our fourth and final technology, \emph{Differential Privacy} (DP) \cite{dwork06}, has the potential to create a separation in time. DP protects against both active and passive attacks. It ensures that the location of any user in the data is hidden when computing statistics by perturbing them before publishing. The perturbation reduces the accuracy of the published statistics, but also obscures the presence of any user in the original data. In addition to its purpose-built use to protect individual privacy, DP has proved useful in anonymizing user location as well \cite{HooffLZZ15,LazarZ16}.

The principal concern pertaining to the application of DP toward our setting is \emph{accuracy}.
Utilizing current DP solutions for infinite streams of cloud utilization data would greatly deteriorate the accuracy of the published statistics by protecting the user whereabouts throughout time.
In the initial works that achieved DP's privacy goals in the streaming setting \cite{dwork10, chan11}, the accuracy of the output greatly deteriorates with the stream size. Fan and Xiong \cite{fan12} show how to publish accurate statistics while satisfying differential privacy, i.e., obscuring the user presence in the data throughout time, assuming a finite stream. Cao et al.~\cite{cao17} also take into account temporal correlations among different locations. On the other hand, Kellaris et al.~\cite{kell14} assume an infinite scheme, but satisfies differential privacy for any time window of a predefined finite size, i.e., the user presence is solely protected within any time interval of fixed length. This final solution offers a potential path forward for us.

\subsection{Symbiosis}

We propose a combination of
the four technologies described above
in order to strike the appropriate balance between security, accuracy, and cost.

We begin by advocating for a datacenter-wide application of MTD. In addition to its benefits for tenants in general, its uncertainty specifically decouples location information in the past vs.~present.

\begin{figure}[t]
\begin{centering}
\includegraphics[width=7cm]{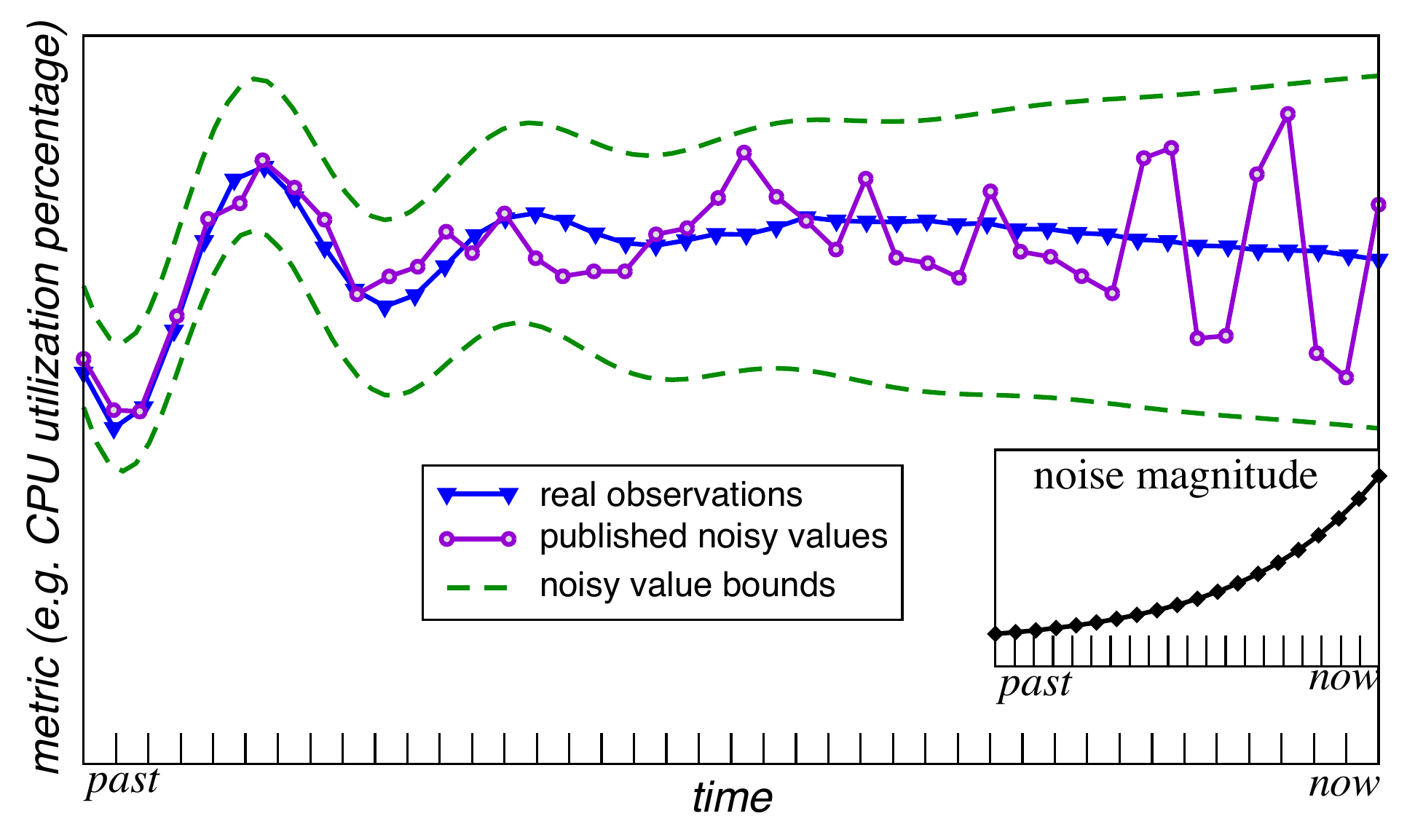}
\caption{A hypothetical CPU utilization graph showing both the real observations (blue) that remain private and the perturbed observations (purple) that we envision releasing publicly. Note that the noise envelope (green) that governs the perturbations varies by time, simultaneously yielding accurate past data for historical analysis and protection against location attacks in the present.}
\label{fig:decayingDifferentialPrivacy}
\end{centering}
\end{figure}

As long as the allocation of nodes within each migration step follows SCS, the attacker is also stymied when trying to execute a co-location attack based solely on side channels available in the present. Next, we advocate that cloud tenants employ PCS so as to ensure that any cryptographic material that may have been compromised in the past is now irrelevant.

Essentially, by periodically applying MTD, we may release statistics about the past to the public. On the other hand, the more recent the statistics, the higher the probability that they incorporate the current location of a user.

As such, we can employ DP in a `decaying' manner to conceal the current whereabouts of the user, while allowing more accurate past. \Cref{fig:decayingDifferentialPrivacy} depicts the `envelope' of uncertainty provided by DP's noise. We highlight the fact that the noise varies over time so that uncertainty rises as one approaches the present.

Combining DP and MTD introduces an inevitable but acceptable trade-off between accuracy and cost. Deploying MTD more frequently decreases the need for publishing inaccurate statistics that satisfy DP; in other words, it permits the `envelope' of uncertainty to decay faster as one goes back through time. However, MTD is resource-intensive, and DP provides security protections for data released more recently.

Future work is needed here because none of the current DP approaches can be combined as-is with the MTD, as they do not take into account the probabilistic nature of the sensitive user location.

As such, we identify the need for a new notion of privacy, which (i)~can be seamlessly combined with MTD, (ii)~quantifies the offered security against co-location attacks, and (iii)~returns useful monitoring statistics with bounded error. Towards this, we plan on viewing MTD as a probabilistic game (similar to \cite{Valizadeh2016MarkovMO}), and defining a new notion of privacy that offers provable guarantees similar to DP, while protecting the locations of the users between executions of the MTD. 



\section{Discussion Topics}

While we espouse a specific vision, our intent in a workshop format is to promote a general discussion of places in which security technologies can enable the introduction of new (non-security-related!)~features that benefit cloud tenants. We believe this is generally an under-tapped area that is ripe for further discussion, which our presentation should promote.
Thus, in this section, we list some of the core open questions that are integral to the continuation of this research. 

\emph{Interest on cloud transparency:}
%
Is there sufficient interest in the data we propose to make public, Would people be willing to pay for more openness in cloud services, whether the benefits can provide a market differentiator if a cloud vendor adopts our vision but its competitors do not, and whether our strategy is viable to implement if the requisite security concerns are adequately addressed.

\emph{Other attack vectors:}
Our vision to increase the transparency of cloud infrastructures introduces a new attack surface that can be exploited in various ways. We discuss co-location attacks and potential remedies using existing security technologies but usage of released data can enable other attack vectors as well. Whether the released data can enable other types of attacks is a matter of discussion.

\emph{Additional foreseeable overheads:} The particular solution we proposed requires rather substantial changes to the behavior of cloud vendors, who must adopt MTD at the datacenter scale, and tenants, who must modify applications they host on the cloud to absorb the side effects of proactive features of the system. Whether there exists other foreseeable overheads is a matter of discussion.



\ifCLASSOPTIONcompsoc
  \section*{Acknowledgments}
\else
  \section*{Acknowledgment}
\fi

This material is based upon work supported by the National Science Foundation under Grants Numbered 1414119, 1347525, 1565387 and 1149232, the MassTech Collaborative Research Matching Grant Program, and the commercial partners of the Massachusetts Open Cloud, which include Brocade, Cisco, Intel, Lenovo, Red Hat and Two Sigma.

{\footnotesize \bibliographystyle{IEEEtran}
\bibliography{main.bib}}

\end{NoHyper}
\end{document}